\documentclass[twocolumn,groupeaddress,superscriptaddress,showpacs,cite,aps,prb]{revtex4-1}
\usepackage{epsfig,pslatex,latexsym,times,amssymb,amsmath,graphicx,bm,revsymb}
\pdfoutput=1
\begin{document}
\author{M. Prada}
\email{mprada@physnet.uni-hamburg.de}
\affiliation{Instituto de Ciencias Materiales de Madrid, ICMM-CSIC, Sor Juana Ines de la Cruz 3, Madrid, Spain}
\affiliation{I. Institut f\"ur Theoretische Physik, Universit\"at Hamburg, Jungiusstr. 9, 20355 Hamburg, Germany}
\author{G. Platero}
\affiliation{Instituto de Ciencias Materiales de Madrid, ICMM-CSIC, Sor Juana Ines de la Cruz 3, Madrid, Spain}
\author{D. Pfannkuche}
\affiliation{I. Institut f\"ur Theoretische Physik, Universit\"at Hamburg, Jungiusstr. 9, 20355 Hamburg, Germany}
\date{\today}
\title{Unidirectional direct current in coupled nanomechanical resonators by tunable symmetry breaking}
\newcommand*{\ud}{\mathrm{d}}
\newcommand*{\ue}{\mathrm{e}}

\begin{abstract}
{
We investigate theoretically the non-linear dynamics of the most fundamental component of a bistable coupled oscillator. 
Under a weak radio frequency excitation, the resonator is parametrically tuned into self-sustained oscillatory regimes.
The  transfer of  electrons from one contact to the other is then mechanically assisted, generating a rectified current.
The direction of the rectified current is, in general, determined by the phase shift between the mechanical oscillations and the signal.
However, we locate intriguing parametrical regions of uni-directional rectified current, resulting from spontaneous
parity-symmetry breaking.
In these regions, a dynamical symmetry breaking is induced by the non-linear coupling of the mechanical and
electrical degrees of freedom.
The achieved unstable regime favors then a uni-directional response, resulting in a direct, positive electric current.
When operating within the Coulomb blockade limit, the charge balance in the oscillators perturbs
drastically the mechanical motion, causing large accelerations that further enhance the shuttle response.
Our results suggest a practical scheme for the realization of  a self-powered device in the nanoscale.
}

\end{abstract}


\maketitle

Bifurcations are closely linked to catastrophes in systems operating in nonlinear regime. 
In general terms, a system experiences a catastrophe when a smooth change in the value of 
a parameter results in a sudden change in the response of the system. 
The response is then termed as parametrically driven. 
Interest in parametrically driven nonlinear dynamics of nano-electromechanical systems (NEMS) 
has grown rapidly over the last few years \cite{villanueva_nano,  midtvedt_nano, requa_apl}. % 
NEMS offer the possibility to realize nanomechanical switches \cite{subramanian_acs}, circuits 
\cite{blicknjop,mahboob,roukesNMC,zhou,palomaki}, %and parametric amplifiers for 
electronic transducers \cite{roukes_nano,bartsch_acs}, 
current rectifiers \cite{pistolesi, ahn}, 
or high-sensitive charge \cite{scheible_apl04}, spin \cite{rugarNature,palyi,chotorlishvili} and
mass sensors \cite{rugarPRL,roukesMD,puller}, as well as the general study of 
nonlinear dynamics of oscillators and resonators \cite{leturceq, vonoppen, drews,faust_nature,faust, cohen}. 
In these systems charge transport is mechanically assisted and typically 
driven by an RF excitation of tunable intensity and frequency. %
When the mechanical degrees of freedom couple to the RF excitation, 
the system may enter a regime of strong nonlinear response as the frequency and the intensity %
of the excitation are slowly varied, hence, parametrically driven.  
Understanding behavior of these devices in the nonlinear regime
can thus point to strategies for engineering self-powered NEMS-based devices \cite{wang}. 

\begin{figure}[!htb]
\includegraphics[width=0.475\textwidth]{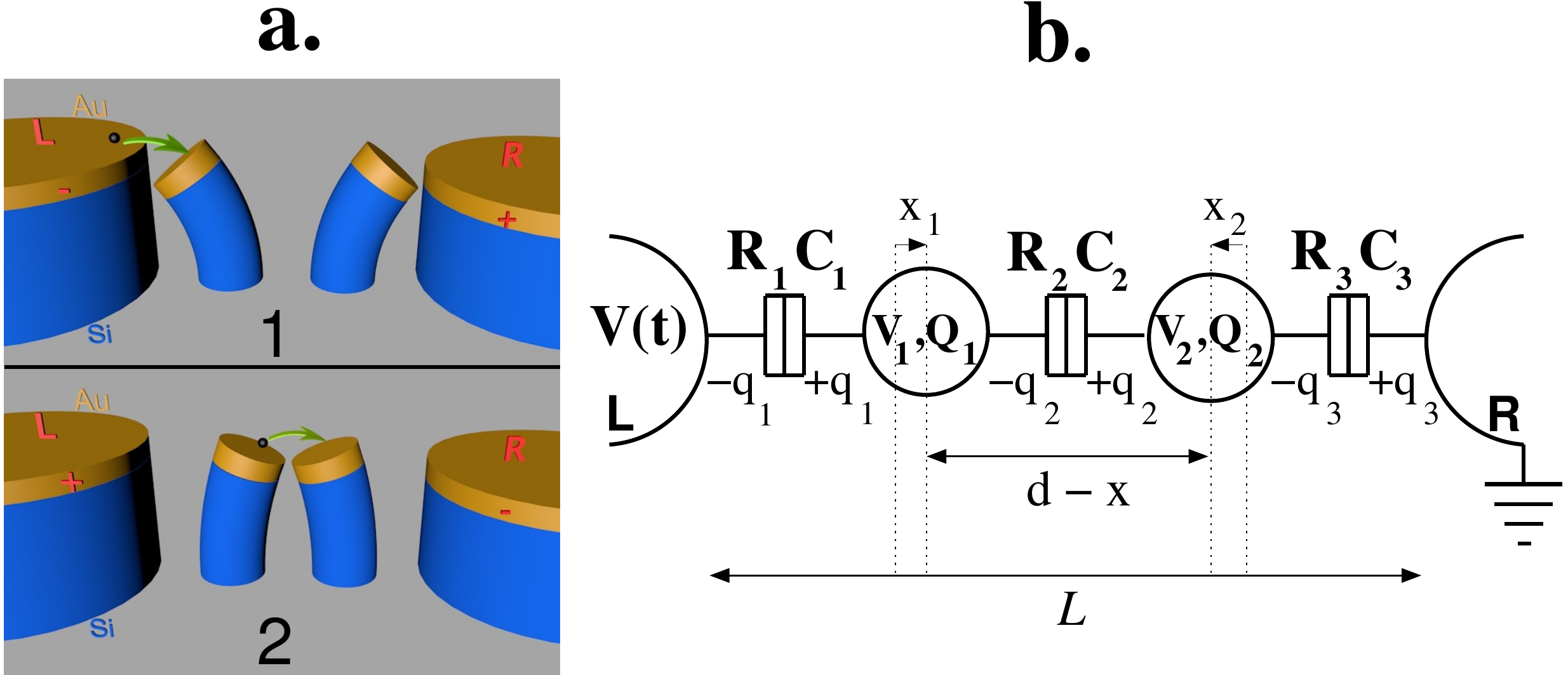}
\caption{ 
a. Sketch showing mechanically assisted electronic transport in Si nanopillars with a metallic island on top. 
The black dot represents the transferred electron. 
The flexural mode which originates an efficient dc current has the center of mass at rest. 
b. Circuit representation: two metallic islands are capacitively coupled to each other and to 
both electrodes, L and R. %, which are connected to an external voltage source, $V(t)$. 
}
\label{fig1}
\end{figure}

Back in 1998 Gorelik {\it et al.} \cite{gorelik}
proposed a nanomechanical transistor based on a vibrating shuttle between two contacts. 
The mechanical motion of the shuttle perturbs the charge balance, causing 
accelerations that could further enhance the response of the shuttle. %an observable current across the system.
Scheible {\it et al.} \cite{scheible_apl04} experimentally realized such an 
electron shuttle in the form of a single nanopillar vibrating in a flexural mode, 
and observed a frequency dependent ratchet behavior.
Following these experimental results,  
Pistolesi {\it et al.} \cite{pistolesi} showed theoretically that 
the single  electron shuttle can indeed act as a rectifier.
More recently, Kim {\it et al.} \cite{ck,ck2,kimprada_prl} realized a 
nanomechanical shuttle capable of transferring electrons mechanically at room temperature 
in the Coulomb blockade regime. 
Their prototype consisted of two coupled Si nanopillars with a metallic island on top, as depicted in  Fig. \ref{fig1}(a). 
The nanopillars are placed between two 
electrodes %(see Fig. \ref{fig1}d) 
operating at high frequencies ($\sim$0.1 GHz). 
One of the great advantages of this system is that cotunneling
events are dramatically suppressed in a flexural oscillatory mode where the 
center of mass is at rest [see  Fig. \ref{fig1}(a)]. 
Another is the obtention of a rectified dc signal resulting from
spontaneous symmetry breaking \cite{ck,ahn}. %, as anticipated by Ahn {\it et al.} \cite{ahn}. 

In this letter, we investigate the dynamics of the most fundamental component for the realization of bistable 
oscillators: a double coupled shuttle NEMS. 
In previous work by Ahn {\it et al.} \cite{ahn}, the symmetry on 
the phase portraits of the solutions prevented a preferred direction for the current.
However, we report here on spontaneous parity-symmetry breaking due to the non-linear coupling 
of the mechanical and electrical degrees of freedom. The parity-symmetry breaking results in the obtention of 
uni-directional direct currents.  
We demonstrate that parametrically controlled current flow and largely amplified response excel the 
double shuttle as an optimal device. 
We explore the parametric excitations within the Coulomb blockade limit, where
abrupt, almost instantaneous increments and decrements of the charge difference in the nano-islands 
cause the average electrostatic force to change periodically in a square-wave fashion.
Corresponding electromechanical instabilities open access to regimes of efficient charge pumping. 

We start with a purely classical description of the circuit depicted in Fig. \ref{fig1}(b), 
as commonly found in literature \cite{pistolesi, ahn,gorelik}. 
The mechanical degree of freedom (see Fig. 1) is described in terms of the 
relative displacement of two islands, $x =x_1 - x_2$. 
The dynamics  within the quasi-adiabatic limit \cite{isacsson,ahn} is governed by 
the force exerted by an average electric field $-V(t)/L$ on the coupled oscillators, yielding
\begin{equation}
\label{eq:dyn}
\ddot{x} + \frac{\gamma}{\omega_0} \dot x+ x= \frac{e V(t)}{kL} n(t),
\end{equation}
where $n(t) = -(Q_1-Q_2)/e$, $Q_{1,2}$ denotes the charge in the nanoislands, 
with $Q_i = q_i-q_{i+1} $,  %$Q_1=q_1-q_2$,  $Q_2 = q_2-q_3$, 
$q_i$ being the charge in the $i$-th junction (see Fig. \ref{fig1}b),   
$k = m\omega_0^2$, with $m$ being the mass of a pillar and
$\omega_0$ the oscillator eigenfrequency, 
$\gamma$ denotes the dynamic damping, % is $\gamma$, 
and $L$ is the distance between the electrodes. 
When the flexural modes of the nanopillars are excited, the %(see Fig. \ref{fig1} b.), the
resistances and mutual capacitances of each junction become sensitive to the 
displacements: $C_i\simeq C_i^0/(1+x_i)$  and  
$R_i=R_i^0\exp{(x_i/\lambda)}$, where $\lambda$ is the phenomenologically introduced tunneling length. 
We take $R_1^0 = R_3^0 = R_2^0/2=R$ and $C_1^0=C_3^0=2C_2^0=C$, 
consistent with previous results \cite{ck2}. 
Classical circuit analysis for the doubly charged shuttle gives 
$V(t) = \sum_i q_i/C_i$, $i = 1,3$,   $q_i/R_iC_i = q_j/R_jC_j$, $j\neq i$, 
allowing us to express the charge on each island %$Q_1=q_1-q_2$ and $Q_2 = q_2-q_3$ 
in terms of the relative displacement $x$ \cite{prada_prb}, 
\begin{equation}
Q_1 =-Q_2\simeq \frac{CV(t)}{2}%\frac{CV(t)}{2}
\left[
\tanh{\frac{3x}{4}}-\frac{\lambda}{d}x e^{\frac{3x}{2}}
\right]
, 
\label{eqQ}
\end{equation}
with $d$ being the  distance at rest between the shuttles.
For small oscillations, $x\lesssim 1$, we expand 
Eq. (\ref{eqQ}) to third order. % in $x$.  %he transformation $t\to\omega_0t$
Eq. (\ref{eq:dyn}) becomes a modified Mathieu equation, 
\begin{equation}
\ddot x + x + \frac{\gamma}{\omega_0} \dot{x} +%\frac{3cV_0^2}{4mL\omega_0^2}
\alpha(\sin{\omega t} + \beta)^2
\left[ x - \frac{2\lambda }{d} x^2- \frac{3}{16} x^3\right]
=0
.
\label{eqX}
\end{equation}
The dimensionless parameter $\alpha = 3CV_0^2/4mL\lambda\omega_0^2$ quantifies the strength of the RF excitation, 
being the ratio of the electric ($\sim CV_0^2/L$) and mechanical forces ($\sim m\lambda\omega_0^2 = k\lambda$), 
with $k$ being the sprig constant. 
$V_0$ and $\beta$ relate to the applied voltage, $V(t) = V_0(\sin{\omega t} + \beta)$. 
Throughout this work we take typical experimental values for $\lambda/d$ = 0.1  and $\gamma/\omega_0$ = 10$^{-2}$. 
Within the weak electromechanical coupling limit, %$1/Q, \alpha \ll $1.
we pa\-ra\-me\-te\-ri\-ze the damping and the excitation strength  
with an arbitrarily small $\epsilon$, $\gamma \sim \epsilon \gamma_1$ and 
$\alpha \sim \epsilon \alpha_1$. We consider the resonant
modes of the system, $\omega \simeq \omega_0(p + \epsilon\delta_\omega)$, 
where $\delta_\omega$ indicates the deviation from the natural resonance 
and $p$ is the winding number \cite{gorelik,ck}.  
This defines two time scales, the ``stretched'' time, 
$z = \omega t$, and the ``slow'' time, $\eta = \epsilon t$.  
We seek for steady oscillatory solutions, 
$x(\eta,z) = A(\eta)\cos{z/p} - B(\eta)\sin{z/p}$. 
Note that the mechanical oscillations can be alternatively expressed as $x(\eta) = r_0(\eta)\cos{[\omega t -\varphi(\eta)]}$, with 
$r_0^2 = A^2 +B^2$, and  
$\varphi = \arctan{B/A}$ defining the phase between the mechanical oscillations and the voltage. 
Following the Poincar\'e-Lindstedt method, 
we find linearized differential equations for the coefficients $A$ and $B$ \cite{prada_prb}.    
Phase portraits and bifurcation diagrams are finally computed using
evaluation routines \cite{matcont_auto}. 

\begin{figure}[!htb]
\includegraphics[width=0.450\textwidth]{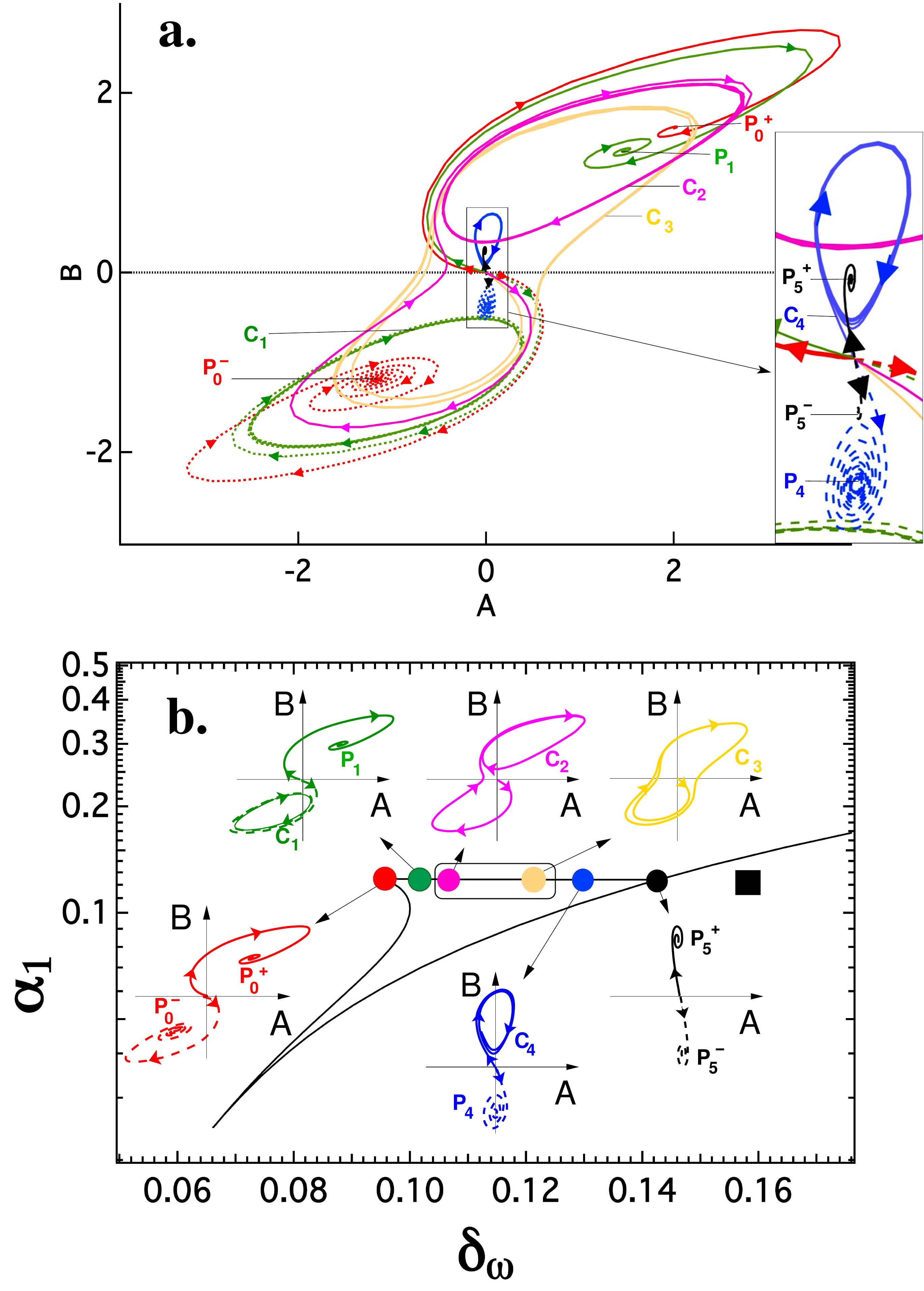}
\caption{ %(color online) 
a. Phase portraits  for the fundamental mode, $p$ = 1. 
Inset: magnification around the origin. 
b. Corresponding bifurcation diagram in parameter space. 
The rectangle marks a band of uni-directional current.
Insets: phase portraits where symmetry breaking occurs.
}
\label{fig2}
\end{figure}

The phase portraits in the first unstable region are shown in Fig. \ref{fig2}(a), with the corresponding 
bifurcation diagram in Fig. \ref{fig2}(b). 
We chose six representing points along the parametric region of multiple stability. 
The associated phase portraits for each point are depicted in the insets [from Fig. \ref{fig2}(a)]. 
For these, the initial conditions were set near equilibrium,  $B_0 = b_0$ ($B_0 = -b_0$) 
for the solid (broken) traces, with arbitrarily small $b_0$. %(red, green, pink and blue) 
The stable solutions are either attractor points, $P_i^\pm$ or cycles, $C_i$, $i$ = 0,$\dots$4. 
This results in electromechanical instabilities, 
in the sense that even if the pillars are initially nearly at rest, 
the electrostatic field will cause them to oscillate. 

A pitchfork bifurcation is observed as we move from the left in the unstable region. 
The origin undergoes a transition, from 
a stable spiral to a saddle, where two non-trivial quasi-symmetric attractors appear (red traces, $P_0^\pm$). 
A supercritical Andronov-Hopf bifurcation causes 
one of the attractors to  become an asymptotic curve ($C_1$).  % disappears under further parametric variations. 
Under further parametric variations a homoclinic bifurcation results in the birth of an unique 
limit cycle ($C_2$, $C_3$). The asymmetry of the cycles result in an unique uni-directional current (see below). 
The rectangle marks the parametric region where uni-directional current is obtained.

The time-average direct current is then obtained for each asymptotic solution ($P_i$, $C_i$) by integrating over a period the 
current across a junction,
\begin{equation}
I_{\rm DC} = \frac{\omega}{4\pi R}\int_{t_0}^{t_0+T}{\rm d}t\frac{V(t)e^{x}}{1+e^{3x/2}\cosh{\left(\frac{X}{2}\right)}}
.
\label{eq:Idc}
\end{equation}
\begin{figure}[!htb]
\includegraphics[width=0.4750\textwidth]{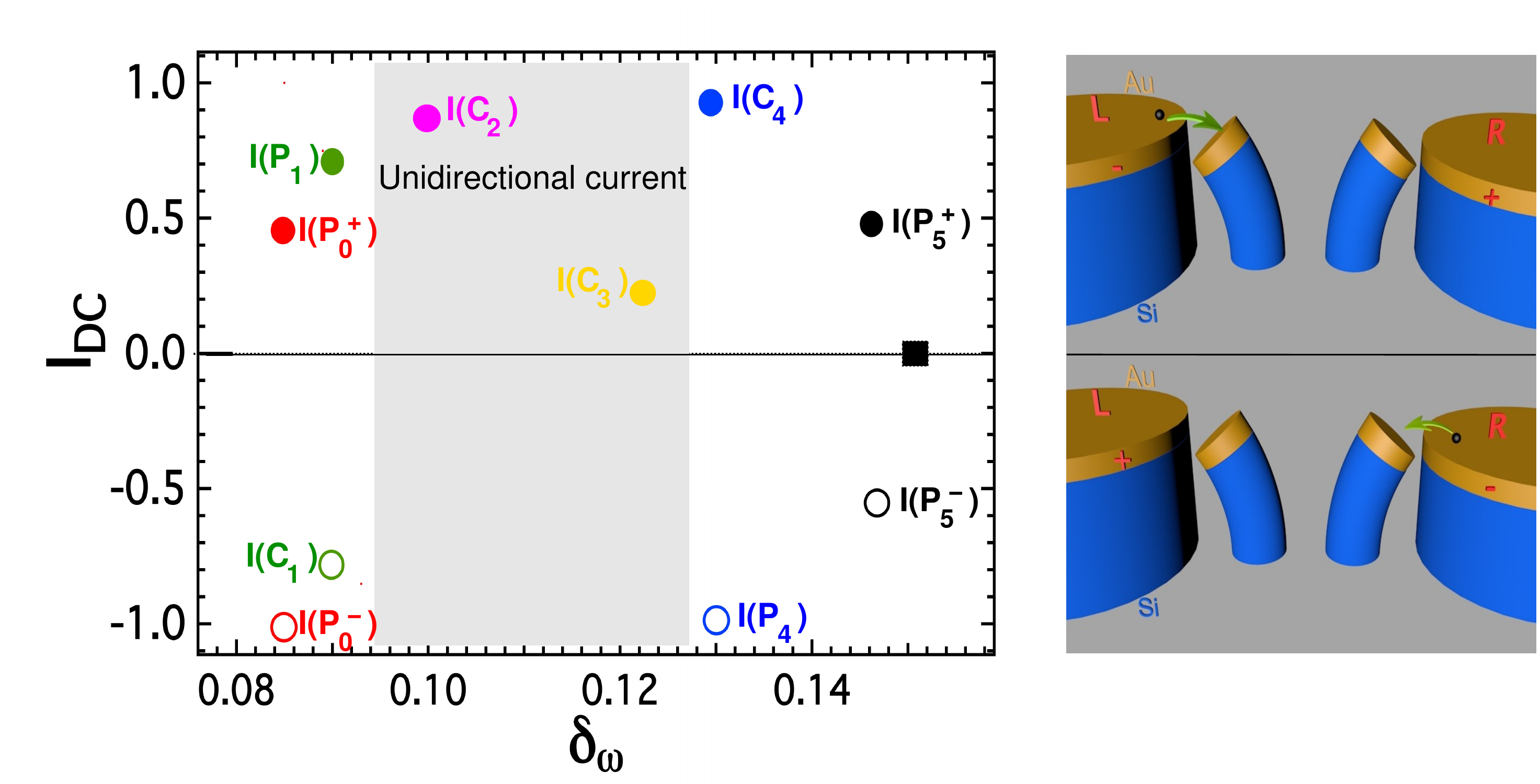}
\caption{ %(color online) 
Computed $I_{\rm{DC}}$ at the stable solutions (arbitrary units). % using the same color coding. 
The solid (open) dots are associated to the solid (broken) trajectories  of the insets of Fig. \ref{fig2}(b) with 
the corresponding color code. In the shaded region the direct current occurs only left-to-right.  
Right insets: sketch of the movement of the pillars. 
For $B>0$ $(<0)$, the mechanical oscillation is in (out of) phase with the applied RF signal 
(see sign on the contacts), and hence, $I_{\rm DC}$ results positive (negative). 
}
\label{fig3}
\end{figure}

Here, $X$ is the center of mass displacement, which in most of the cases remains at rest  \cite{ahn}. 
The results are summarized in Fig. \ref{fig3}. 
An intuitive picture is sketched in the right insets of Fig. \ref{fig3}: 
for $B>0$, the mechanical oscillations are always in phase with the voltage and electrons are
flowing from left to right. % contact in all cases. 
However, for $B<0$, the oscillators are out of phase, resulting in a negative rectified current. 
Outside the unstable region [black square of Fig. \ref{fig2}(b)] the 
trivial solution ($A$ = $B$ = 0) is an attractor, and no direct current is observed. 
The sign of the current in typical bistable regions is then determined by the initial conditions, $\pm b_0$.
This occurs in the red, green, blue and black traces of Fig. \ref{fig2}(a) where the sign of the phase is preserved, or, in other words, 
the trajectory remains in one semiplane ($B>0$ or $B<0$). 
The corresponding stroboscopic plot is nearly symmetric in these bistable regions, so a positive or negative current can occur. 
On the contrary, for the magenta and yellow traces of Fig. \ref{fig2}(a), 
only one stable asymptotic orbit is found ($C_2$ and $C_3$). 
The lack of symmetry of the stable phase portrait result in a positive (left-to-right) net current {\em independently} of the initial conditions. 
This occurs throughout the shaded region of Fig. \ref{fig3}.
%In the shaded region,thus,  the asymmetry of the phase portraits result in a left to right net current. 
We note that the term in $x^2$ of Eq. (\ref{eqX}), breaks the parity symmetry of the system: 
for a given solution $x(t)$, $-x(t)$ is not longer a solution. 
This term results from relative  variation of the individual capacitances and resistances with the positions 
of the islands. 
Hence, we conclude that  a dynamical symmetry breaking occurs due to the non-linear coupling of the mechanical 
and electrical degrees of freedom. 
We stress that the sign of the current is then parametrically controlled, suggesting 
an energy harvester for self-powered nanosystems \cite{li_n}, or a {\em nano-battery}. 

%Within this region, two non-trivial stable nodes coexist in the phase space,
%one with $B<0$ and another with $B>0$.  % [{\it eg.} red or black curves in Fig. \ref{fig2}(a)]. 
%These points become Hopf bifurcations under further variations of $\alpha$, 
%while the origin remains unstable. % spiral.  

\begin{figure}[!htb]
\includegraphics[width=0.4750\textwidth]{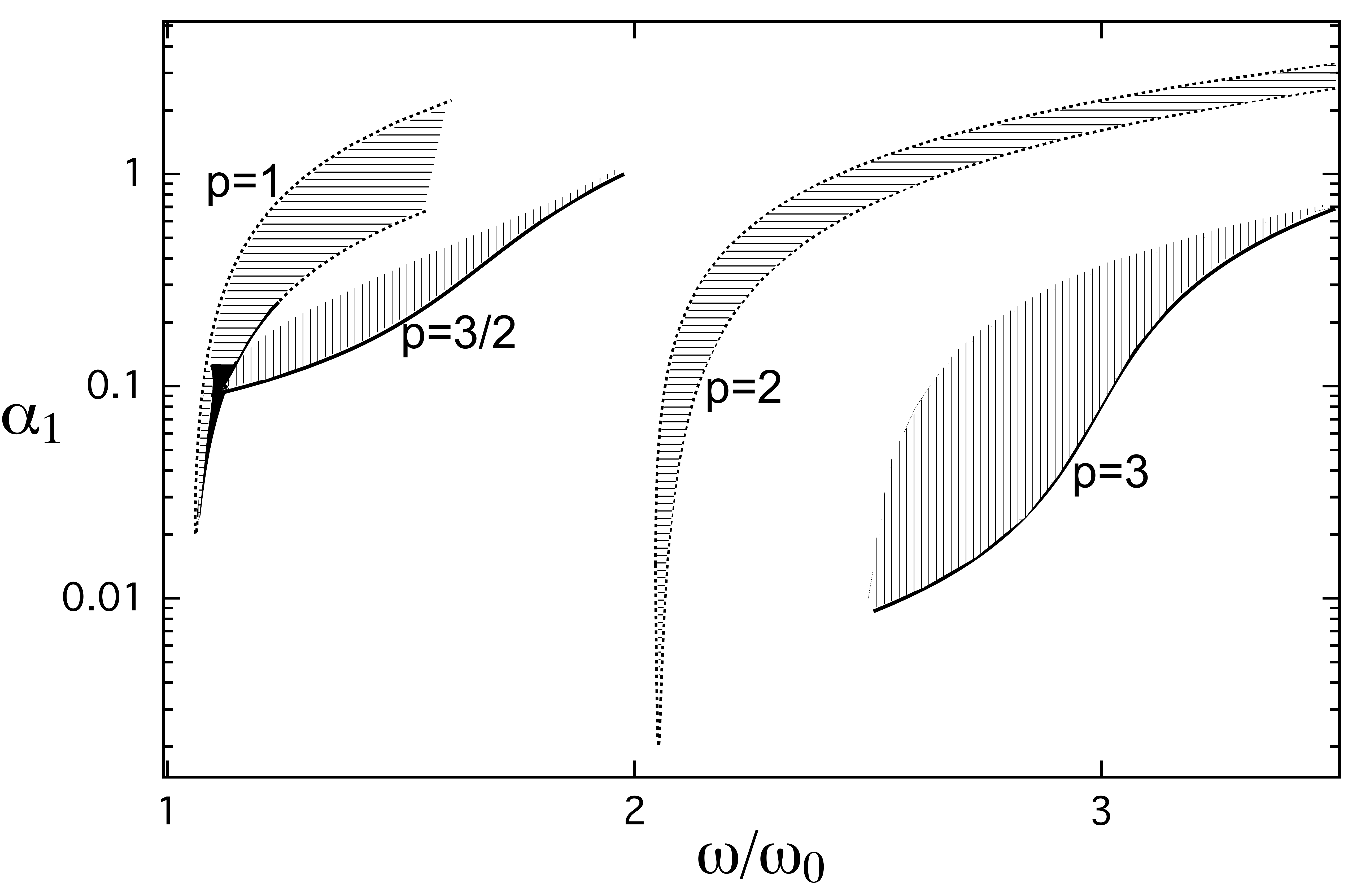}
\vspace{-0.2 cm}
\caption{
Bifurcation diagram showing regions with multiple stability in the areas
where the origin is unstable (horizontally hatched, $p$ = 1,2) or stable 
(vertically hatched, $p$ = 3/2, 3). 
The dark region in $p$ = 1 corresponds to Fig. \ref{fig2}(b). 
}
\label{fig4}
\end{figure}

We now study the response on the relevant 
higher excitation frequency regions, {\it i.e.}, $p=3/2,2,3$ in Eq. (4). %\ref{Eq:ABDC}. 
For these, non-trivial solutions are found in the hatched 
areas indicated in Fig. \ref{fig4}. 
The case $p=2$  is of particular interest, as there exists a region in 
parameter space for which the origin becomes unstable. %saddle node. 
We then have a situation similar to the one of Fig. \ref{fig2}, within a different region of 
parameter space (see Fig. \ref{fig3}). 
For the cases $p = 3/2, 3$, the origin of the phase plane is always an attractor, 
and hence, solutions with initial conditions in its proximity will vanish.  
However, another three attractors exist around a small area of the phase plane. 
%similar to the one marked in the right-top corner inset of Fig. \ref{fig2} a. 
Only when the initial conditions are close to these attractors, self-sustained 
oscillations are possible for this mode, but no electromechanical instabilities are found.

Next, we evaluate the dynamics of the islands within the Coulomb blockade limit, 
for which $CV\simeq e$. 
In such a picture, abrupt increments and decrements of 
charge in the metallic islands occur almost instantaneously, separated by periodic time intervals \cite{weiss_zwerger}. 
We use the master equation in terms of the excess electrons in the islands, $n_{1,2}$, 
\begin{equation}
\label{eq:mastereq}
\dot P_{n_1,n_2} = \sum_{n_1,n_2;k}\Gamma^k_{n_1^\prime,n_2^\prime\to n_1,n_2}P_{n_1^\prime,n_2^\prime} -
\Gamma^k_{n_1,n_2\to n_1^\prime,n_2^\prime}P_{n_1,n_2},
\end{equation}
where the tunneling rate at the $k^{\mathrm{th}}$ junction is given, according to the orthodox theory, by: 
$\Gamma^k_{i\to j} =   \mu_{ij}(t)/e^2R_k(t)[1-e^{-\mu_{ij}(t)/k_BT}]$, % and $\gamma^k_0(t) = 1/e^2R_k(t)$. %, $k$ = 1-3.
with $\mu_{ij}$ being the decrease of free energy  when the tunneling event occurs. 
We solve Eq. (\ref{eq:mastereq}) by direct integration and get the occupation on each 
island, $\langle n_i(t)\rangle$. 
\begin{figure*}[!htb]
\includegraphics[width=0.75\textwidth]{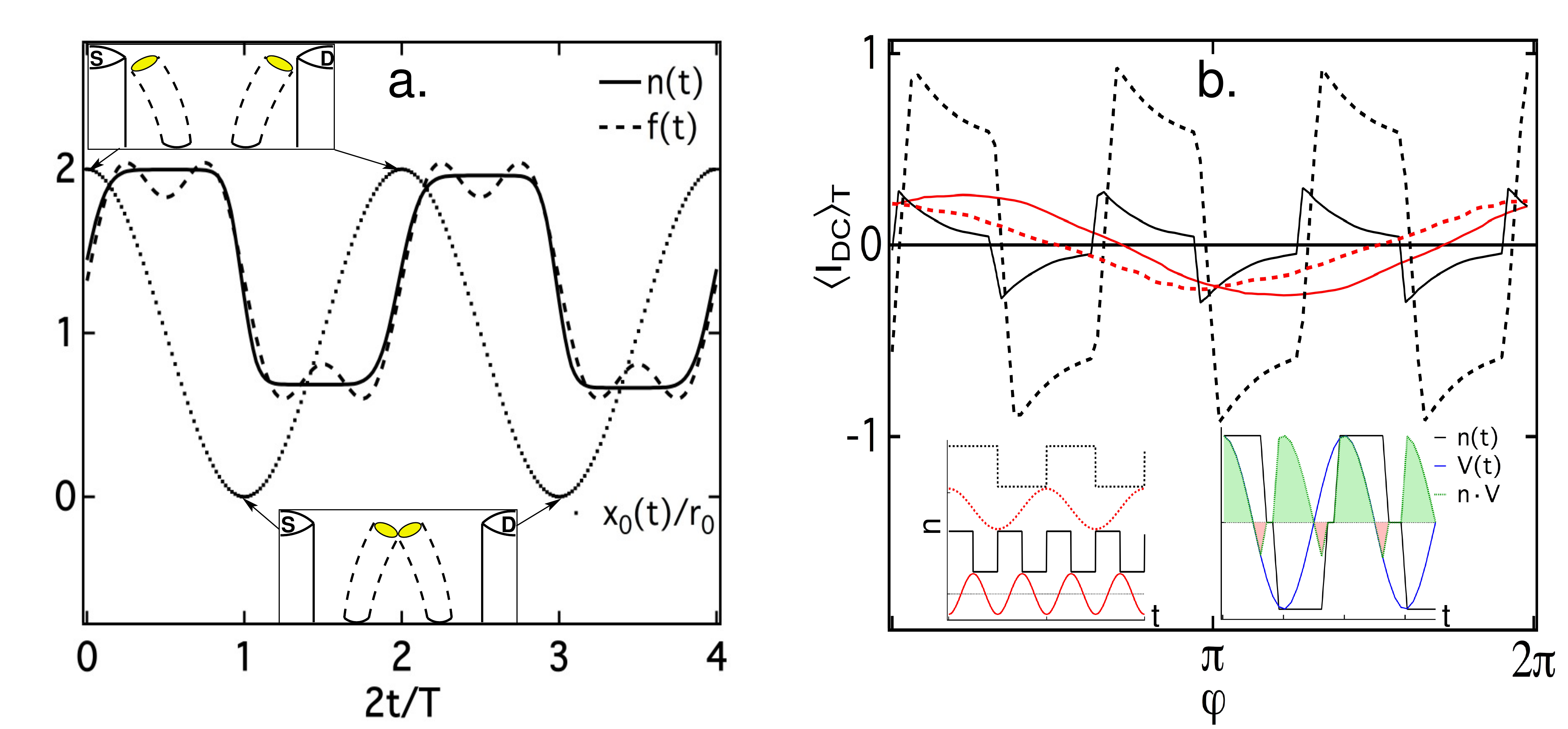}
\caption{
a. $n(t)$ calculated using the master equation 
(Solid curve), square wave fit, $f(t) = n_{\mathrm{av}} + 4n_0(\cos{\omega_0t} - \cos{3\omega_0t})/\pi$
(broken curve) and normalized relative displacement, $x_0(t)/r_0$ (dotted curve) . 
The insets mark the points of maximal displacement for the pillars.
b. Averaged dc current per period for four different charge oscillations. %Inset: amplitude of oscillations as a function of $\alpha^\prime_1$. 
Lower-left inset: The corresponding charge oscillations. 
Lower-right inset: $n(t)$ (solid black curve) and RF signal (blue trace). 
The resulting direct current per period is then proportional to the green shaded areas minus the pink ones. 
}
\label{fig5}
\end{figure*}
Fig. \ref{fig5}(a) shows the steady state solution for $\langle n(t)\rangle$. 
As it can be seen, charge transfer occurs in the points of maximal deflection. 
The relative charge is nearly a square wave, 
$\langle n(t)\rangle \sim n_{\mathrm{av}} + 4n_0(\cos{\omega_0t} - \cos{3\omega_0t})/\pi$. 
Inserting this expression into Eq. (\ref{eq:dyn}), we  find, as before, linearized 
equations for the coefficients $A$ and $B$ of the oscillatory solutions, 
\begin{eqnarray}
\label{eq:AB_CB}
2\frac{\ud A }{\ud \eta} &= &
-{A}{\gamma_1} -  \frac{2\delta_\omega}{p}  B
+ \alpha_1^\prime n_{\mathrm{av}}\delta_{p,1} - \frac{\alpha^\prime_1}{6}(2\delta_{p,2} + \delta_{p,4})
%\nonumber \\ && 
\nonumber \\
%\cos{z}\to
2 \frac{\ud B }{\ud \eta} &= &
-{B}{\gamma_1} +  \frac{2\delta_\omega}{p} A
%\nonumber \\ && 
+ \alpha^\prime\beta n_0, 
\end{eqnarray}
with $\alpha^\prime = eV_0/L k \lambda$ and 
$\alpha^\prime_1  = \epsilon \alpha^\prime$. 
It is straight forward to see that %, for $\beta\neq 0$ or $p$ = 1, 2 or 4, \ref{eq:AB_CB} 
Eq. (\ref{eq:AB_CB}) has one non-trivial attractor in the phase plane: 
In the absence of $V_{{\mathrm{DC}}}$ ($\beta = 0$), stable points are found
only around subharmonics with $p$ = 1, 2 or 4,  
($A_p$, $B_p$) = $a_p(\gamma, \gamma_p)$,
with $\gamma_p = 2\delta_\omega/p$, $a_{1} = \alpha^\prime n_{\mathrm{av}}/(\gamma^2 + \gamma_1^2)$,
$a_{2} = -\alpha^\prime n_{0}/3(\gamma^2 + \gamma_2^2)$
and $a_{4} = -\alpha^\prime n_{0}/6(\gamma^2 + \gamma_4^2)$.
%will occur in the catastrophic regions marked in Fig. \ref{fig3}. 
In contrast, under a finite dc bias, the left-right symmetry  is broken, and oscillatory solutions 
are found for {\em any} frequency at  ($A_p$, $B_p$) = $\alpha\beta n_0(\gamma_p, -\gamma)/(\gamma^2+\gamma_p^2)$.
Let us focus now on the existence of a self-sustained oscillation. 
In an oscillatory mode, $x\sim r_0 \cos{(\omega_0t + \varphi)}$, 
the absorbed power by the oscillator per unit cycle is obtained 
by averaging over a period the last term of Eq. (\ref{eq:dyn}):
\[
\langle W_{\mathrm{a}}\rangle_{} \simeq \left\langle {\alpha^\prime n(t)\dot x(t)(\sin{\omega t} + \beta)}\right\rangle, 
\]
whereas the dissipated power is given by the damping term, 
$\langle W_{\mathrm{dis}}\rangle_{} = \langle \dot x^2\rangle\gamma/\omega_0$. 
If this amount is larger than the dissipated power, 
$\langle W_{\mathrm{a}}\rangle_{} \gtrsim \langle W_{\mathrm{dis}}\rangle_{}$, 
self-sustained oscillations \cite{kim_njp10} are expected. This occurs
when max$\{\alpha^\prime n_{\mathrm{av},0}, \alpha^\prime n_0\beta\} \gtrsim r_0\gamma/\omega_0$, 
with the appropriate phase $\varphi$. % \cite{unpublished2}.  
The amplitude of the oscillations could then be large enough to reach the Fowler-Nordheim tunneling limit \cite{scheible_prl}, 
with a subsequent enhancement of the direct current due to field emission. % \cite{ck3}.
Hence, under a finite dc bias, bands of instabilities occur for a particular value of the applied RF power, $\alpha$. 
Recent experimental data confirm the existence of such bands \cite{kimprada_prl}.

To illustrate the importance of the phase between the mechanical oscillations and the RF signal, 
we compute the current for a few fictitious charge oscillations $n(t)$.
Fig. \ref{fig5}(b) shows the average direct current (number of transferred electrons per cycle)
as a function of the phase $\varphi = \arctan{\{B/A\}}$ 
for four different charge oscillations, sketched in the lower-left inset. 
The function have the same period as the signal (solid traces) or twice (broken traces), 
and is a square wave (black) or a sinusoidal (red). 
It is evident from the figure that the 
square wave function with a periodicity twice the signal one (broken black traces) shows 
the most effective current on a broad relative phase range.  
A square-wave fashion variation is indeed
known to pump energy more effectively than a sinusoidal variation \cite{butikov}. 
This results in a rapid and enhanced response of the system, 
a very feature desired for industrial applications.  
The current is then negative for $(2n-1)\pi/3 \lesssim \varphi\lesssim 2n\pi/3$ and positive elsewhere. 
We stress that the unidirectional current flow translates from the classical limit to the Coulomb blockade regime. 
The lower-right inset represents the RF signal (blue) and the corresponding  $n(t)$ (black). 
The averaged current per period is then proportional to the shaded areas, which is, for this particular 
choice of phase ($\varphi$=0), non-zero and positive.  

In summary, we have theoretically studied a weakly AC driven coupled electron shuttle.
We find multiple stability regions in parameter space with subsequent %
self-sustained oscillations generating a finite observable direct current.
In the bi-stable regions, the sign of the current is bound to the relative phase of the mechanical oscillators 
and the signal. % 
Subsequent dynamical symmetry breaking result in the obtention of uni-directional direct currents, 
a key feature in the implementation of a nano-battery. 
Within the Coulomb blockade regime, electromechanical instabilities are observed, which occur at any 
frequency in the presence of a dc bias. 
The tunability of the self-sustained oscillations and the uni-directionality of the current
suggest a vast number of potential applications. 

{\bf  Acknowledgments:}  
We are grateful to R. H. Blick and C. Kim for enlightening discussions.
This work was supported by the  program SB2009-0071, 
MAT 2011-24331, ITN grant 234970  and GrK 1286.

%\bibliography{refs}
%\end{document}

\end{document}